\documentclass[aps,twocolumn,prx,superscriptaddress]{revtex4-1}
\newcommand{\pagenumbaa}{1}
\usepackage{graphicx}
\usepackage{amsmath}
\usepackage{siunitx}
\usepackage{xcolor}
\usepackage{soul}
\definecolor{prxblue}{rgb}{0,0,0.57}
\usepackage[colorlinks=true,
            linkcolor=prxblue,
            urlcolor=prxblue,
            citecolor=prxblue]{hyperref}
\newcommand{\equ}[1]{Eq.\,(\ref{#1})}
\newcommand{\Equ}[1]{Equation\,(\ref{#1})}
\newcommand{\fig}[1]{Fig.\,\ref{#1}}

\newcommand{\be}{\begin{equation}}
\newcommand{\ee}{\end{equation}}
\newcommand{\bea}{\begin{eqnarray}}
\newcommand{\eea}{\end{eqnarray}}

\newcommand{\mb}{\mathbf}

\newcommand{\mbf}{\mathbf}

\begin{document}

\title{Symmetry Breaking of the Persistent Spin Helix in Quantum Transport}

\author{Pirmin~J.~Weigele}\altaffiliation{These authors contributed equally to this work.}
\affiliation{Department of Physics, University of Basel, CH-4056, Basel, Switzerland}

\author{D.~C.~Marinescu}\altaffiliation{These authors contributed equally to this work.}
\affiliation{Department of Physics and Astronomy, Clemson University, Clemson, South Carolina 29634, USA}

\author{Florian~Dettwiler}
\affiliation{Department of Physics, University of Basel, CH-4056, Basel, Switzerland}

\author{Jiyong~Fu}
\affiliation{Instituto de F\'{\i}sica, Universidade de Bras\'{\i}lia, Bras\'{\i}lia-DF 70919-970, Brazil}

\author{Shawn~Mack}
\affiliation{U.S. Naval Research Laboratory, Washington, DC 20375, USA}

\author{J.~Carlos~Egues}
\affiliation{Instituto de F\'isica de S\~ao Carlos, Universidade de S\~ao Paulo, 13560-970 S\~ao Carlos, S\~ao Paulo, Brazil}

\author{David~D.~Awschalom}
\affiliation{Institute for Molecular Engineering, University of Chicago, Chicago, Illinois 60637, USA}

\author{Dominik~M.~Zumb\"uhl}
\affiliation{Department of Physics, University of Basel, CH-4056, Basel, Switzerland}

\date{\today}
\begin{abstract}
We exploit the high-symmetry spin state obtained for equal Rashba and linear Dresselhaus interactions to derive a closed-form expression for the weak localization magnetoconductivity -- the paradigmatic signature of spin-orbit coupling in quantum transport. The small parameter of the theory is the deviation from the symmetry state introduced by the mismatch of the linear terms and by the cubic Dresselhaus term. In this regime, we perform quantum transport experiments in GaAs quantum wells. Top and back gates allow independent tuning of the Rashba and Dresselhaus terms in order to explore the broken-symmetry regime where the formula applies. We present a reliable two-step method to extract all parameters from fits to the new expression, obtaining excellent agreement with recent experiments. This provides experimental confirmation of the new theory, and advances spin-orbit coupling towards a powerful resource in emerging quantum technologies.

\end{abstract}

\maketitle

\setcounter{page}{\pagenumbaa}
\thispagestyle{plain}

\section{Introduction}
The spin-orbit (SO) interaction is of profound importance for a broad range of phenomena in modern condensed matter physics, such as spin textures~\cite{Schliemann2017,Kohda2017}, spin Hall effects~\cite{Kato2004,Sinova2004}, topological insulators~\cite{Bernevig2006a,Konig2007,Liu2008,Knez2011} and Majorana fermions~\cite{Lutchyn2010,Oreg2010}, as well as for application in spintronics~\cite{Zutic2004,Awschalom2011} and quantum computation \cite{Loss1998,Hanson2007,Kloeffel2013}. Semiconductors such as GaAs, InAs, or GaSb offer various strengths of SO coupling combined with a high level of electrical control~\cite{Engels1997,Nitta1997,Papadakis1999,Grundler2000,Koga2002,Kohda2012,Yoshizumi2016,Dettwiler2017} over the SO parameters e.g. in quantum wells and are thus suitable for a broad range of experiments. The two dominant contributions to SO coupling in semiconductor quantum wells arise from breaking of structural and bulk inversion symmetry, quantified by the Rashba coefficient $\alpha$ and the Dresselhaus coefficient $\gamma$, respectively. While the Rashba effect~\cite{Rashba1984} is linear in electron momentum, the bulk Dresselhaus~\cite{Dresselhaus1955} term is cubic. When projected into a quantized 2D system, it retains a cubic component with coefficient $\beta_3$ but also acquires a linear component of strength~$\beta$.

A particularly interesting situation arises when $\alpha=\beta$: a persistent spin helix (PSH) can be formed~\cite{Schliemann2003,Bernevig2006}, which is robust against D'yakonov Perel scattering, strongly suppressing spin relaxation~\cite{Koralek2009}. In this state, spins do not precess at all when traveling ballistically along one particular direction in the 2D plane, while precessing quickly when proceeding along the orthogonal direction in the 2D plane. Effectively, spin symmetry is restored by a complete cancellation of the Rashba and linear Dresselhaus terms along one direction and the creation of a uniaxial internal SO field -- broken only by the cubic Dresselhaus term and by a deviation from $\alpha=\beta$.

With optical methods, the SO parameters can be readily extracted from experiments~\cite{Kalevich1990,Kato2004a,Meier2007,Studer2009,Koralek2009,Eldridge2011,Walser2012a,Ishihara2013,Kunihashi2016} by monitoring a spin component directly e.g. with Kerr rotation methods. This is much more difficult to achieve from electronic transport measurements where the spin information is not usually directly accessible. In materials with strong SO coupling, the beating patterns of the Shubnikov de-Haas oscillations can sometimes be used to extract the Rashba parameter~\cite{Das1989,Engels1997,Nitta1997,Papadakis1999,Grundler2000}. Even if SO coupling is weak, quantum interference effects depend very sensitively on the spin of the electron, giving weak antilocalization (WAL) as the paradigmatic signature of SO coupling in quantum transport experiments. To extract the SO parameters from such highly-sensitive magnetoconductance measurements, one needs to rely on a model containing the relevant SO terms. For some special cases, it was possible to derive closed-form expressions already early-on: with cubic terms only \cite{Hikami1980,Altshuler1981}, without SO terms altogether \cite{Hikami1980,Altshuler1980}, or for the spin helix point $\beta=\pm\alpha$ and $\beta_3=0$, in which case weak localization (WL) was recovered \cite{Pikus1995} as if there were no SO coupling at all. It is clear that these are very isolated special cases of limited practical use.

A closed-form expression for the quantum corrections to the magnetoconductance that incorporates all the SO coupling terms identified above is highly desirable not just for its fundamental theoretical value, but particularly also for applications, where it is important to be able to extract the SO parameters from transport data in order to control and engineer devices. This is required to turn SO coupling into a powerful resource for quantum technologies. Moreover, a closed-form theory is clearly preferable over a numerical expression which can be difficult or unpractical to handle for data fitting. However, despite almost 4 decades of considerable efforts, it was not possible to provide such a closed-form expression containing cubic and both linear terms.

The spin helix state -- predicted \cite{Schliemann2003,Bernevig2006} and realized \cite{Koralek2009,Walser2012,Kohda2012} only relatively recently -- affords a new opportunity in tackling this long-standing and unresolved problem by offering a high symmetry point around which a new small parameter may be introduced: the deviation from the perfect spin symmetry, via imperfectly matched linear terms $\propto \alpha-\beta$ or through the cubic term $\propto \beta_3$. In both cases of broken spin symmetry, the effective SO field remains small, i.e. the WAL minima occur at small magnetic field, thus remaining in the spin diffusive regime.

Here, we exploit this new small parameter and are able to derive a new closed-form expression including cubic and both linear SO terms in the vicinity of the PSH point by following the established WAL formalism. Further, we develop a reliable method to extract all relevant SO parameters from quantum transport data using the new expression. This method exploits the cancellation of the linear terms to first extract independently the cubic term and phase coherence in the high carrier density regime where the cubic term already breaks spin symmetry and restores WAL. Then, we tune slightly away from $\alpha=\beta$ and can now also extract the linear SO parameters, again from fits to the new theory. This two stage procedure delivers all SO parameters, in very good agreement with recent transport studies~\cite{Krich2007,Dettwiler2017} as well as optical experiments~\cite{Walser2012a,English2013,Altmann2016}. In particular, we extract a Dresselhaus material parameter $\gamma=11.5\pm1\,\mathrm{eV\AA^3}$ in good agreement with recent experiments.

\section{Theory of Quantum Corrections to Conductivity}
There is a large body of literature addressing the subject of quantum corrections over the past decades: already the very early work of Hikami, Larkin and Nagaoka \cite{Hikami1980} includes SO effects in the form of impurity scattering (skew scattering) in the diffusive regime and is the only work to date to provide a closed-form expression in presence of SO interaction and a magnetic field. The effect of an in-plane magnetic field was also discussed soon after ~\cite{Maekawa1981}. For the case of III-V semiconductors where the Dyakonov-Perel mechanism \cite{Dyakonov1986} is prevalent in the diffusive regime, the linear and cubic Dresselhaus terms were included in presence of a magnetic field \cite{Iordanskii1994}, providing an analytical but not closed-form expression. A similar expression is obtained when only the Rashba term is retained~\cite{Punnoose2006}. For the generic case with both Rashba and linear as well as cubic Dresselhaus terms, a closed-form or analytical expression is not available and a numerical solution has to be obtained \cite{Pikus1995,Knap1996}.

Beyond the diffusive regime, only skew scattering was considered \cite{Kawabata1984,Zduniak1997} and had to be solved numerically. Both Rashba and Dresselhaus terms could be treated but only numerically and without taking into account coherent interference effects between the terms \cite{Miller2003}. More complete numerical models exist for either only Rashba or only linear Dresselhaus terms~\cite{Golub2005} or also for  all three terms ~\cite{Glazov2006,Glazov2009,Sawada}.

Here, we consider a 2D electron gas placed in the $\hat{x}-\hat{z}$ plane and the $\hat{y}$-axis perpendicular to the plane. The single particle Hamiltonian corresponding to an electron of effective mass $m^\ast$, momentum $\mathbf{p}=\{p_x,p_y,p_z\}$ and spin $\sigma=\{\sigma_x,\sigma_y,\sigma_z\}$ with Rasbha and Dresselhaus SO coupling reads
\begin{equation}
\begin{split}
H_\mathbf{p}&=\frac{p_x^2+p_z^2}{2m^\ast}+\alpha(\sigma_zp_x-\sigma_xp_z)+\beta_1(\sigma_zp_z-\sigma_xp_x)\\
&-\gamma(\sigma_zp_zp_x^2-\sigma_xp_xp_z^2)\;,
\end{split}
\end{equation}
where $\beta_1$ is the bare linear Dresselhaus coefficient. This choice of coordinates highlights the existence of a $\hat{z}$ in-plane axis, obtained through a $\pi/4$ in-plane rotation to be parallel to $[1\bar{1}0]$ ($\hat{x}\parallel [110]$), that becomes the quantization axis for the electron spin. At $\alpha = \beta$ the spin projection on this axis is a good quantum number of the system, a property not immediately apparent if one chooses the standard designation of $\hat{z}$ perpendicular on the plane.

Since the conduction in the degenerate Fermi system is realized only by states at the Fermi surface of wave vector $k_F$, $p_x$ and $p_z$ are expressed as a function of the polar angle $\varphi_\mathbf{p}$ between the momentum $\mathbf{p}$ and the $[110]$ axis. In this case the Dresselhaus Hamiltonian obtains two distinct angular symmetries, effectively renormalizing the linear Dresselhaus strength to $\beta$~\cite{Iordanskii1994,Pikus1995,Dettwiler2017}. We can now write the single particle Hamiltonian in terms of symmetric (+) and antisymmetric (-) combinations of the linear SO couplings, as
\begin{equation}
H_\mathbf{p}=\frac{\mathbf{p}^2}{2m^\ast}+\hbar(\mathbf{\Omega}_\mathbf{p}\times \sigma)\cdot \hat{y}.
\label{eq:sohamil}
\end{equation}
The SO coupling is expressed via $\mathbf{\Omega_p}$, which is defined as
\begin{align}
\hbar\Omega_\mathbf{p}^x&=k_F\left[(\alpha+\beta)\cos\varphi_{\mathbf{p}} - \beta_3 \cos 3\varphi_{\mathbf{p}}\right]\;,\\
\hbar\Omega_\mathbf{p}^z&=k_F\left[(\alpha-\beta)\sin\varphi_{\mathbf{p}} - \beta_3 \sin 3\varphi_{\mathbf{p}}\right]\;,
\end{align}
where $\beta=\beta_1-\beta_3$ is the renormalized linear Dresselhaus coefficient. We follow the standard formalism to calculate the quantum corrections to the conductivity~\cite{Iordanskii1994,Pikus1995,Altshuler1980,Rammer1998} for the single particle Hamiltonian in \equ{eq:sohamil}.

The quantum corrections to the conductivity result from the renormalization of the scattering matrix element through the coherent superposition of the incident and scattered states. Although the bare impurity scattering is considered to be spin-independent, in the presence of spin-orbit coupling, an additional spin component is involved in the calculated effective value of the matrix element. This is a result of the slight change in the energy of the electrons when the backscattered momentum is not perfectly anti-parallel, but rather deviates by a small vector $\mb q$. The ensuing variation in energy $\Delta E (\mb q)$, considered small when compared with the energy uncertainty in the collision process $\hbar/\tau_0$, depends simultaneously on the two spin states of the electrons before and after the collision, which are considered uncorrelated. In a perturbative approach that involves a power expansion in $\Delta E(\mb q)\tau_0/\hbar$, the renormalization is done through the eigenvalues of an operator, called the Cooperon, acting in the 4-dimensional space associated with the two spin 1/2 particles. The eigenvalues of this operator then yield the corrections to the conductivity when summed over all the changes $\mb q$ and spin channels. Here, in the vicinity of the spin helix symmetry, these eigenvalues can be calculated exactly and we obtain a closed-form expression for the quantum corrections.

The possible total-spin states formed correspond either to total angular momentum $J=0$, the singlet $S$, or to the total angular momentum $J=1$, the triplet states $T_0$ and $T_\pm$, labeled after the values of $J_z=0,\pm 1$. The associated four eigenvalues make up the quantum corrections in a system with SO coupling. The singlet is antisymmetric under the exchange of the incident and scattered spins, leading to an additional minus sign, thereby making the singlet contribution positive and, thus, responsible for the antilocalization contribution to the conductivity. The triplet states, on the other hand, are all symmetric and contribute negatively to the conductivity upon backscattering, thus making up the localization contribution to the conductivity.

\begin{figure}
\includegraphics[width=8.6cm]{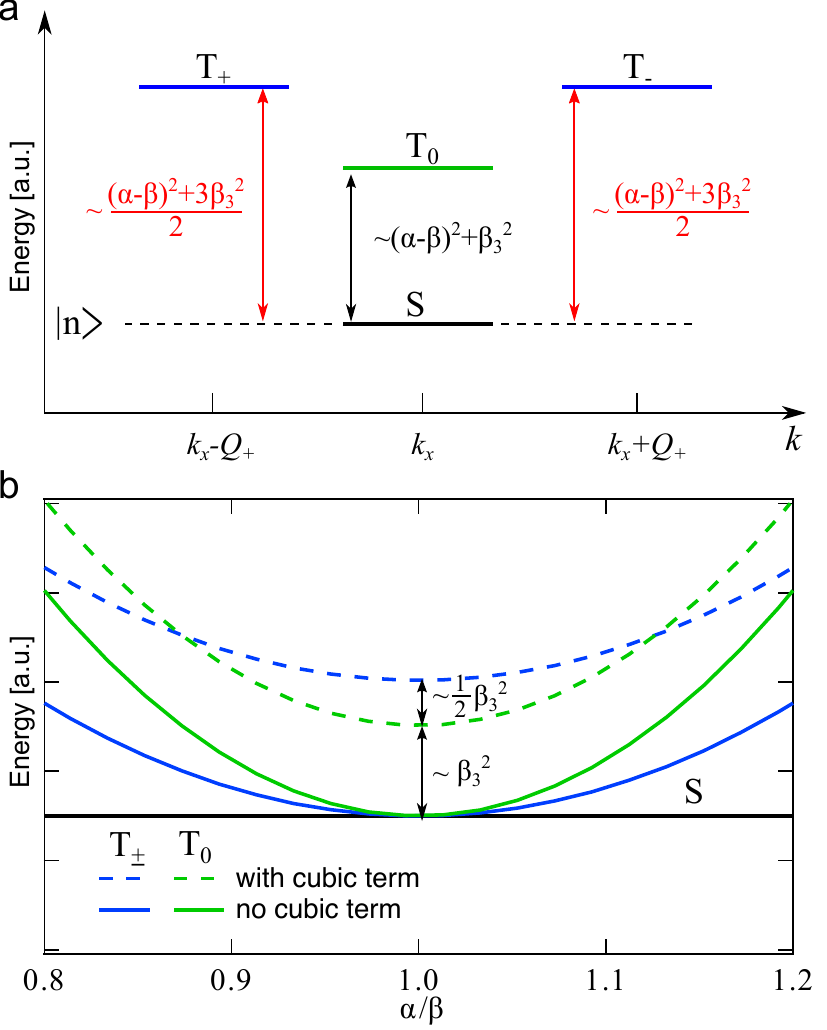}
\vspace{-6mm}\caption{Cooperon terms around the PSH symmetry $\alpha = \beta$, with singlet ($S$) and triplet ($T_0$, $T_\pm$) states in a generic Landau level $|n\rangle$. (a) Energy of the Cooperon eigenstates as functions of $k_x$, the Cooperon momentum along $\hat{x}$ which fixes the center of the orbit. The $S$ and $T_0$ states are located at $k_x$ and become degenerate at $\alpha=\beta$ and $\beta_3=0$. $T_+$ and $T_-$ are degenerate, but since the orbits are separated by $2Q_+$ there is no coupling between them, giving WL. (b) Energies of the eigenstates in one Landau level as a function of the ratio of $\alpha/\beta$. The full curves correspond to the case where the cubic term is zero and all states are degenerate at $\alpha=\beta$ and WL is observed. The dashed lines correspond to the states, when the cubic term is strong, then $S$ and $T_0$ state are not degenerate at $\alpha=\beta$ giving WAL even at $\alpha=\beta$.}
\label{fig:LL}\vspace{-4mm}
\end{figure}

If a magnetic field is applied, the electron energy is quantized in Landau levels (LL) of index $n$. In this case, the magnetoconductivity corrections are evaluated from a properly normalized sum that incorporates all the spin channels in all LL. The interplay between the Landau level quantization and the action of the SO coupling in determining the WL contribution in the $\alpha = \beta$ regime is illustrated in \fig{fig:LL}(a). For any given Landau level $|n\rangle$, we plot the energy of the orbit with respect to the singlet state and indicate the values of the Cooperon wave vector $k_x$ along $\hat{x}$, which fixes the center of the orbit.

As we show in the Appendix, when $\alpha \simeq \beta$, the coupling between the triplet modes decreases so much that it can be considered independent in a first order approximation. This is a consequence of the electron spins becoming polarized along the $\hat{z}$ direction under the action of an effective magnetic field $\sim (\alpha + \beta)$, an orientation that is left unchanged  by the scattering process. In the vicinity of this high spin-symmetry point, the orbits of the triplet states are all separated in momentum space with $T_+$ located at $k_x - Q_+$, $T_0$ at $k_x$ and $T_-$ located at $k_x+Q_+$, where $Q_+=\frac{2m^\ast}{\hbar^2}(\alpha+\beta)$. The energy of the orbits with the $T_\pm$ states is proportional to $((\alpha-\beta)^2+3\beta_3^2)/2$, while that of the state $T_0$ is proportional to $(\alpha-\beta)^2+\beta_3^2$, as shown in \fig{fig:LL}(a). The four associated eigenstates are written in the tensor product space between the LL representation and the total angular momentum representation as $|n\rangle \bigotimes |J,J_z\rangle$. The corresponding Cooperon wave vector $k_x$ is introduced in the position representation of $|n\rangle$.

Although the energies of the parallel spin modes $T_\pm$ are equal, the misalignment along $k_x$ with the center orbits separated by exactly $2Q_+$ precludes any coupling between these modes. This situation corresponds to the separation in the momentum space of the two Fermi populations of up and down spin electrons by $Q_+$, that become spin polarized by an effective magnetic field proportional to $(\alpha + \beta)$~\cite{Bernevig2006}. (The Cooperon is composed of two electrons, so the single particle states are separated in the momentum space by  $Q_+$.) The remaining modes with $J_z = 0$, whose orbits are located at $k_x$, generate opposite sign contributions to WL. Exactly at $\alpha=\beta$ and $\beta_3=0$ they cancel, leading to the disappearance of the WAL. In \fig{fig:LL}(b) we illustrate how the states in the same Landau level evolve as a function of $\alpha/\beta$ for zero cubic term (full curves) and finite cubic term (dashed curves), which highlights the role played by the cubic Dresselhaus term, lifting the degeneracy at $\alpha=\beta$ such that the $T_0$ and $S$ state no longer fully cancel, giving WAL even at $\alpha=\beta$.

In the Appendix we outline the major steps for the calculation (with further details in the SM) while here we give only the result of the closed-form expression for the conductivity correction $\Delta\sigma(B_\bot)$ in a magnetic field $B_\bot$, expressed in terms of the digamma function $\Psi$,
\begin{equation}
\begin{split}
\Delta \sigma(B_\bot)=&-\frac{e^2}{4\pi^2\hbar}\left[\Psi\left(\frac{1}{2}+\frac{B_\varphi}{B_\bot}\right)+2\ln\frac{B_{\text{tr}}}{B_\bot} \right.\\
&-2\Psi \left(\frac{1}{2}+\frac{B_\varphi}{B_\bot}+\frac{B_{SO-}+3B_{SO3}}{2B_\bot}\right)\\
&\left.-\Psi\left(\frac{1}{2}+\frac{B_\varphi}{B_\bot}+\frac{B_{SO-}+B_{SO3}}{B_\bot}\right)\right].
\label{eq:dsigma}
\end{split}
\end{equation}
The coherence time $\tau_\varphi$ and transport time $\tau_\mathrm{tr}$ define two characteristic fields, the dephasing field $B_\varphi$ and the transport field $B_\mathrm{tr}$, which are given by
\begin{subequations}
\begin{align}
B_{\varphi}=\frac{\hbar}{4eD\tau_\varphi}, \label{eq:bphi}\\
B_{\mathrm{tr}}=\frac{\hbar}{4eD\tau_{\mathrm{tr}}}, \label{eq:btr}
\end{align}
\end{subequations}
with $D$ the diffusion constant in 2D.

The form of \equ{eq:dsigma} is very similar to the one from Hikami, Larkin and Nagaoka~\cite{Hikami1980}, but now the arguments in the digamma functions contain the linear Rashba and Dresselhaus terms as well as the cubic Dresselhaus term, via the effective magnetic fields $B_{\mathrm{SO-}}$ and $B_{\mathrm{SO3}}$. These are defined as
\begin{subequations}
\begin{align}
B_{\mathrm{SO}\pm}=&\frac{\hbar}{4e}\left(\frac{2m^\ast}{\hbar^2}(\alpha\pm\beta)\right)^2, \label{eq:bso2} \\
B_{\mathrm{SO3}}=&\frac{\hbar}{4e}\left(\frac{2m^\ast}{\hbar^2}\beta_3\sqrt{\frac{\tau_3}{\tau_1}}\right)^2, \label{eq:bso3}
\end{align}
\end{subequations}
where $\hbar$ the reduced Planck constant and $e$ the elementary charge. The contribution of the cubic Dresselhaus term $\beta_3$ is represented in \Equ{eq:bso3}, multiplied by the square root of the ratio of the backscattering time $\tau_1$ and its third harmonic $\tau_3$ which arises due to the higher angular harmonics of the Dresselhaus term in the SO Hamiltonian~\cite{Iordanskii1994,Pikus1995} (see Eq.~(S4) in SM). In modulation doped structures, the doping layer is set back from the 2D electron gas. Compared to doping incorporated inside the quantum well, this creates a softer, longer range scattering potential for the electrons with more prevalent small angle scattering~\cite{DasSarma1985,Coleridge1991}. For the ratio of scattering times, the range of possible values is $1/9\leq \tau_3/\tau_1 \leq 1$, where $1/9$ corresponds to dominant small angle scattering~\cite{Knap1996} and $1$ indicates short range scattering (isotropic).
\Equ{eq:dsigma} is valid in the diffusive regime, where $B_\mathrm{tr}\ll B_\bot$ and naturally requires weak SO coupling. This is assuming that the spins are precessing only by a small angle in a time $\tau_\mathrm{tr}$, corresponding to the condition $B_{\mathrm{SO}\pm}\ll B_\mathrm{tr}$.

\begin{figure}
\includegraphics[width=8.6cm]{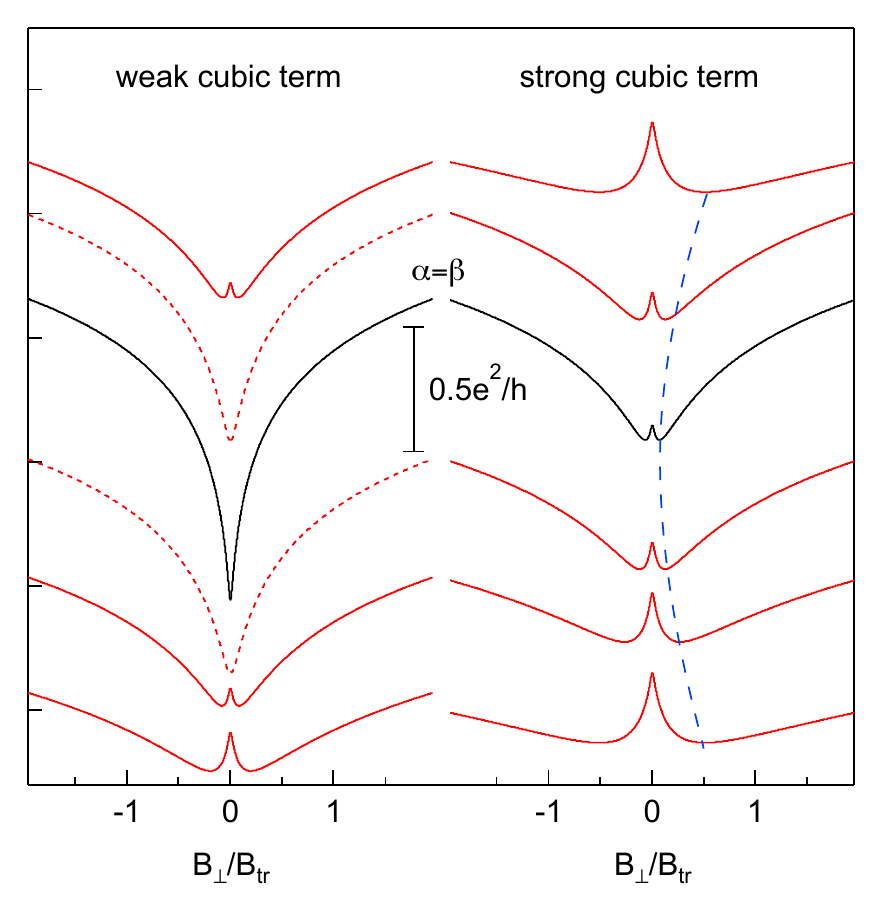}
\vspace{-9mm}\caption{Magnetoconductance curves in the regime close to the spin helix symmetry as given by \equ{eq:dsigma}. The black traces correspond to $\alpha=\beta$. Left panel: Spin orbit coupling causes a quench of the WL before WAL appears (dashed red traces). Right panel: For a strong cubic term, WAL appears even at $\alpha=\beta$ and is defined by the WAL trace with the WAL minima closest to $B_\bot=0$}
\label{fig:0}\vspace{-2mm}
\end{figure}

In \fig{fig:0} we plot the magnetoconductance according to \equ{eq:dsigma} with and without the cubic Dresselhaus term. As we vary the Rashba strength $\alpha$ while keeping the renormalized Dresselhaus term $\beta$ constant, the conductivity traces transition from WAL (red traces) to WL (black trace), where $\alpha=\beta$. We note that the absence of WAL alone (red dashed and black traces, left panel) does not uniquely identify the PSH symmetry point. Rather, the most pronounced WL curve (black trace) with the deepest and sharpest dip indicates realization of the PSH point. Some small amount of SO coupling (cubic and/or linear terms) away from the symmetry point quenches WL, reducing the depth and sharpness of the WL dip without the appearance of WAL, i.e. a maximum of conductivity at zero field. A lower coherence time has a similar effect, also reducing the depth of the WL dip, and can be difficult to separate from the effects of weak SO coupling~\cite{Zumbuhl2002,Zumbuhl2004,Zumbuhl2005}. If a sufficiently strong cubic term is present, WL is suppressed and WAL appears even at $\alpha=\beta$ (black trace), where the position of the WAL minima (indicated by the dashed blue curve) are closest to $B_\bot=0$.

\section{Experiment}
\subsection{Control of Spin Orbit Parameters}
We will now discuss the different constituents of \equ{eq:bso2} and \equ{eq:bso3} and how they relate to experimental adjustable parameters. Electric fields, doping and the interface of the heterojunction result in a confining potential, which causes structure inversion asymmetry and is the origin of the Rashba effect~\cite{Rashba1984}. Its strength $\alpha$ can be tuned as a function of the electric field~\cite{Engels1997,Nitta1997} and is parameterized in our QW as follows
\begin{equation}
\alpha=\alpha_0+\alpha_1\delta E_z,
\label{eq:alpha}
\end{equation}
where $\alpha_0$ is a sample specific offset and $\alpha_1$ accounts for the effect of the induced electric field detuning $\delta E_z$ coming from the voltages applied to the top and back gates (see \equ{eq:detuning} in the Appendix). The Dresselhaus SO interaction \cite{Dresselhaus1955} is characterized by the renormalized linear Dresselhaus strength $\beta$, which reads
\begin{equation}
\beta=\beta_1-\beta_3=\gamma\left(\langle k_z^2\rangle-\frac{k_F^2}{4}\right),
\label{eq:beta}
\end{equation}
where $\beta_1=\gamma\langle k_z^2 \rangle$ and $\beta_3=\frac{1}{4}\gamma k_F^2$ is the cubic Dresselhaus term, with $\gamma$ being the bulk Dresselhaus material coefficient. As the Fermi momentum $k_F^2=2\pi n$ depends on the density $n$, the renormalized Dresselhaus strength becomes controllable via gate voltages, which has recently been demonstrated~\cite{Dettwiler2017}. Over the range of the applied gate voltages $\langle k_z^2 \rangle$ is effectively constant.

\subsection{Evaluation procedure}
\begin{figure}
\includegraphics[width=8.6cm]{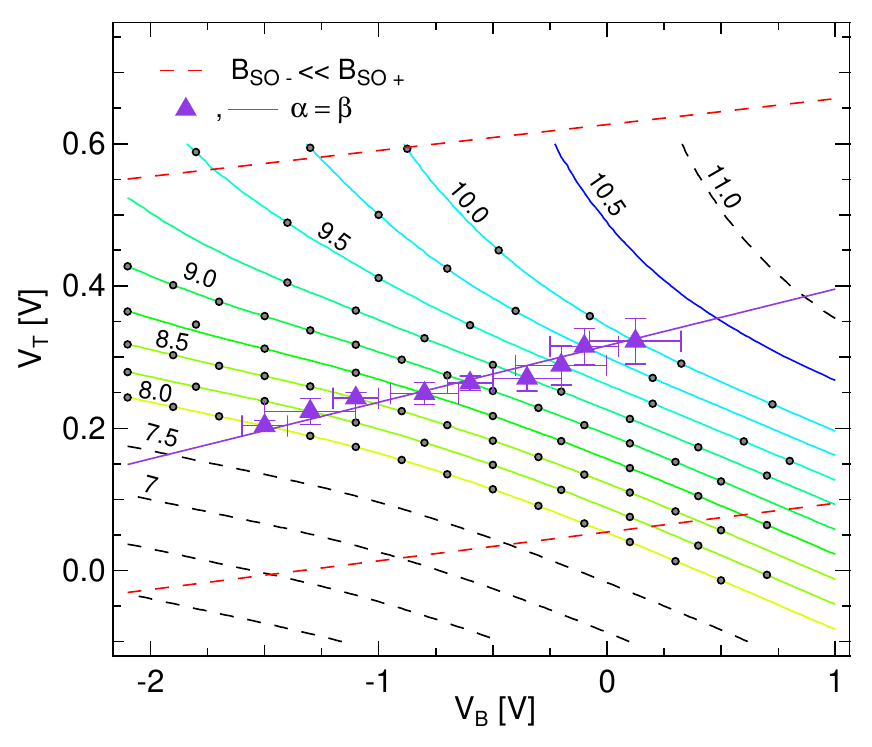}
\vspace{-7mm}\caption{Density map with symmetry points (purple triangles) as a function of top $V_T$ and back gate $V_B$ voltage for data set \#2 (see supplementary for details). Along contours of constant density (labeled in units of 10$^{15}$m$^{-2}$), $B_{SO-}$ is changing as a function of detuning $\delta E_z$, while $B_{SO3}$ is constant. The gray circles indicate the measured gate configurations. The triangles correspond to the approximate position where $\alpha=\beta$ and the purple line corresponds to a plot of the calculated PSH condition from the extracted SO parameters. \equ{eq:dsigma} is valid everywhere between the red dashed lines.}
\label{fig:1}\vspace{-3mm}
\end{figure}

In the experiment, we first extract the cubic term and phase coherence where the linear terms cancel but the cubic term already breaks spin symmetry. Then, we detune the linear terms away from equal size and can extract their strength as well, again from fits to the new theory. We control the strength of the SO parameters $\alpha$ and $\beta$ with the top gate voltage $V_T$ and the back gate voltage $V_B$. As described in the previous paragraphs, these parameters depend on density $n$ and the detuning $\delta E_z$. To obtain a more useful parameter space, we measure the density as a function of $V_T$ and $V_B$ and obtain a density map, shown in \fig{fig:1}, with contours of constant density, along which the detuning $\delta E_z$ changes. We note that for sufficiently negative back gate and positive top gate voltages, the contours of constant density become non-linear, which limits the usable range of $\delta E_z$ and $n$. The range of the density is further limited by the requirement that the cubic Dresselhaus term $\beta_3$, which depends on density, be large enough, such that $B_{\mathrm{SO3}}$ causes WAL even at the PSH symmetry.

The PSH symmetry points are indicated by the purple markers in \fig{fig:1} and their position is estimated from the conductivity traces with the least pronounced WAL feature. This can be done, since along contours of constant density only $B_{\mathrm{SO}-}$ changes as a function of $\delta E_z$ and $B_{\mathrm{SO3}}$ remains constant, as scattering potentials do not change $\tau_3/\tau_1$ significantly for constant density. The gate configurations where conductivity traces were measured are indicated by the gray circles in \fig{fig:1}. At the gate configurations around the symmetry point, $B_{\mathrm{SO}-}$ is very small and is set to zero when fitting \equ{eq:dsigma} to the data, where only $B_{\varphi}$ and $B_{\mathrm{SO3}}$ are the fit parameters (see Appendix Sec. D and Supplemental Material Sec. III.A). The transport field $B_{\text{tr}}$ is known from independent Hall measurements of density $n$ and mobility $\mu$. Since the symmetry point is not precisely known, we determine $B_{\mathrm{SO3}}$ very similarly at the surrounding gate configurations and take the average value, thus obtaining a more robust value for $B_{\mathrm{SO3}}$.

\begin{figure}
\includegraphics[width=8.6cm]{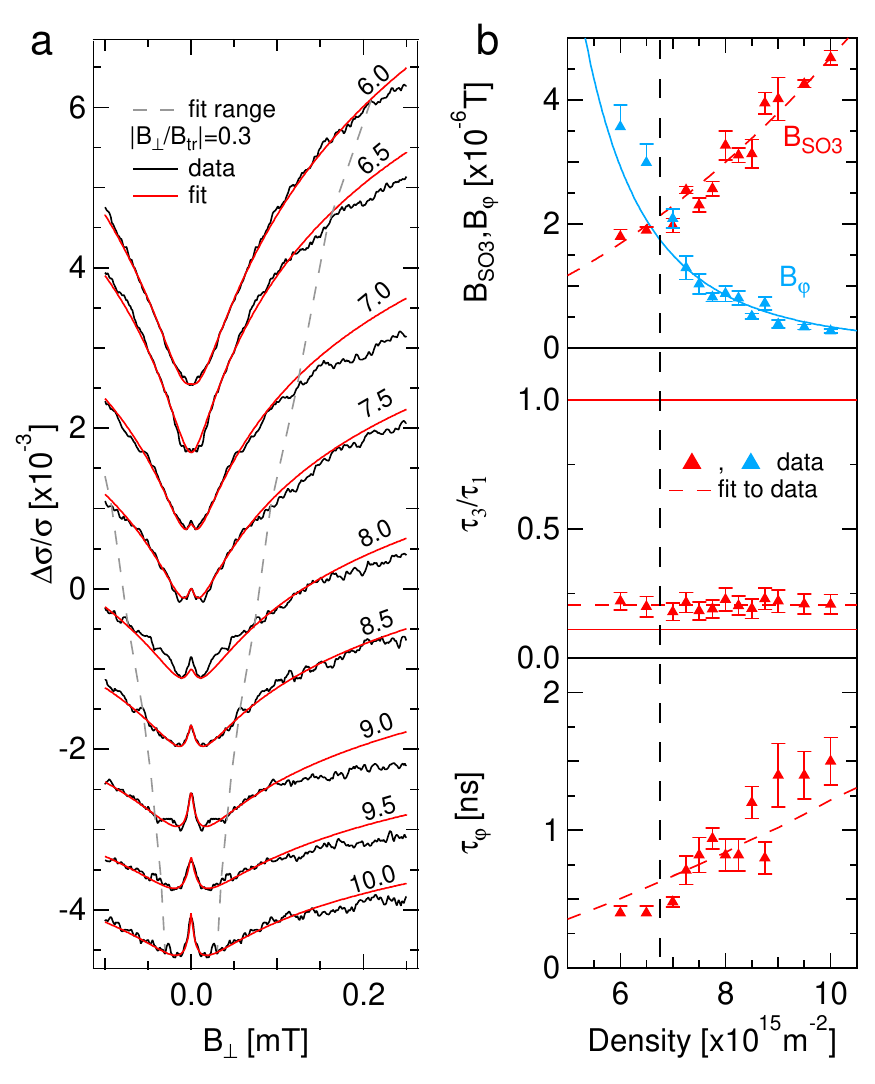}
\vspace{-7mm}\caption{(a) Measured conductivity traces (black) around the symmetry point and fits (red) using \equ{eq:dsigma} with the respective density labeled at each trace in units of 10$^{15}$m$^{-2}$. The measured traces have been symmetrized in $B_{\bot}$ for fitting. The gray dashed curves correspond to the fit range obeying $B_{\bot}\ll B_{\text{tr}}$. (b) \textit{Upper panel}: extracted $B_{\mathrm{SO3}}$ values versus density. The red dashed curve is a quadratic fit to $B_{\mathrm{SO3}}$. The blue markers and curve correspond to the extracted and calculated $B_\varphi$ (see main text). \textit{Middle panel}: extracted $\tau_3/\tau_1$ using the later to be determined $\gamma$ for each individual value of $B_{\mathrm{SO3}}$ from the upper panel. The red dashed line is the average of $\tau_3/\tau_1$. \textit{Lower panel}: coherence time from the extracted $B_{\varphi}$ for the respective density and mobility. The red dashed curve is a fit to the data assuming Nyquist dephasing. For the two lowest densities 6.0 and 6.5 (left of the dashed vertical line), the extracted values of $B_{\mathrm{SO3}}$ and $B_\varphi$ are only bounds, see text.}
\label{fig:2}\vspace{-3mm}
\end{figure}

In \fig{fig:2}(a) we show typical fits (red) to the measured (black) conductivity traces around the symmetry point. The agreement between fit and theory is very good for $B_\bot\ll B_{\text{tr}}$, where $B_{\text{tr}}$ is indicated by the dashed gray curve. The extracted fit parameters $B_{\mathrm{SO3}}$ (red triangles) and $B_{\varphi}$ (blue triangles) are shown as a function of density in the upper panel of \fig{fig:2}(b). A quadratic fit (see \equ{eq:bso3}) to the $B_{\mathrm{SO3}}$ data finds good agreement, see red dashed line. At low temperatures, Nyquist dephasing dominates~\cite{Altshuler1985} and $\tau_\varphi^{-1}\propto T \lambda_F/l_e$, with $T$ being the electron temperature, $\lambda_F$ the Fermi wavelength and $l_e$ the mean free path. Here, the electron temperature is $\sim$100~mK estimated independently~\cite{Casparis2012,Maradan2014}. Since $B_\varphi\propto \tau_\varphi^{-1}$, we can express $B_\varphi$ in terms of density and mobility via the above expression for $\tau_\varphi$. This is shown with the blue curve, reproducing the trend of the extracted $B_\varphi$ quite well. For $n<7\times$10$^{15}$cm$^{-2}$, indicated by the dashed black line in \fig{fig:2}\,b), we observe that the conductivity traces in \fig{fig:2}\,a) no longer show a WAL feature and that $B_{\mathrm{SO3}} \leq B_\varphi$. Thus for densities to the left of the black dashed line, the extraction of a meaningful value for $B_{\mathrm{SO3}}$ and $B_{\varphi}$ is no longer possible and only an upper bound can be determined.

Using the value of $B_{\varphi}$ we can also determine the coherence time $\tau_\varphi$ for each density, which is shown in the lower panel of \fig{fig:2}(b). The coherence time is of the order of 1\,ns, which is a value expected in GaAs 2D electron gases at mK temperatures~\cite{Huibers1998,Huibers1999,Miller2003}. The red dashed curve shows the dependence of $\tau_\varphi$ on density, calculated also for Nyquist dephasing, in qualitative agreement with the data. This allows us to keep $\tau_\varphi$ constant along contours of constant density as the mobility change of $\sim$10\,\% is smaller than the error on $\tau_\varphi$.

\begin{figure}
\includegraphics[width=8.6cm]{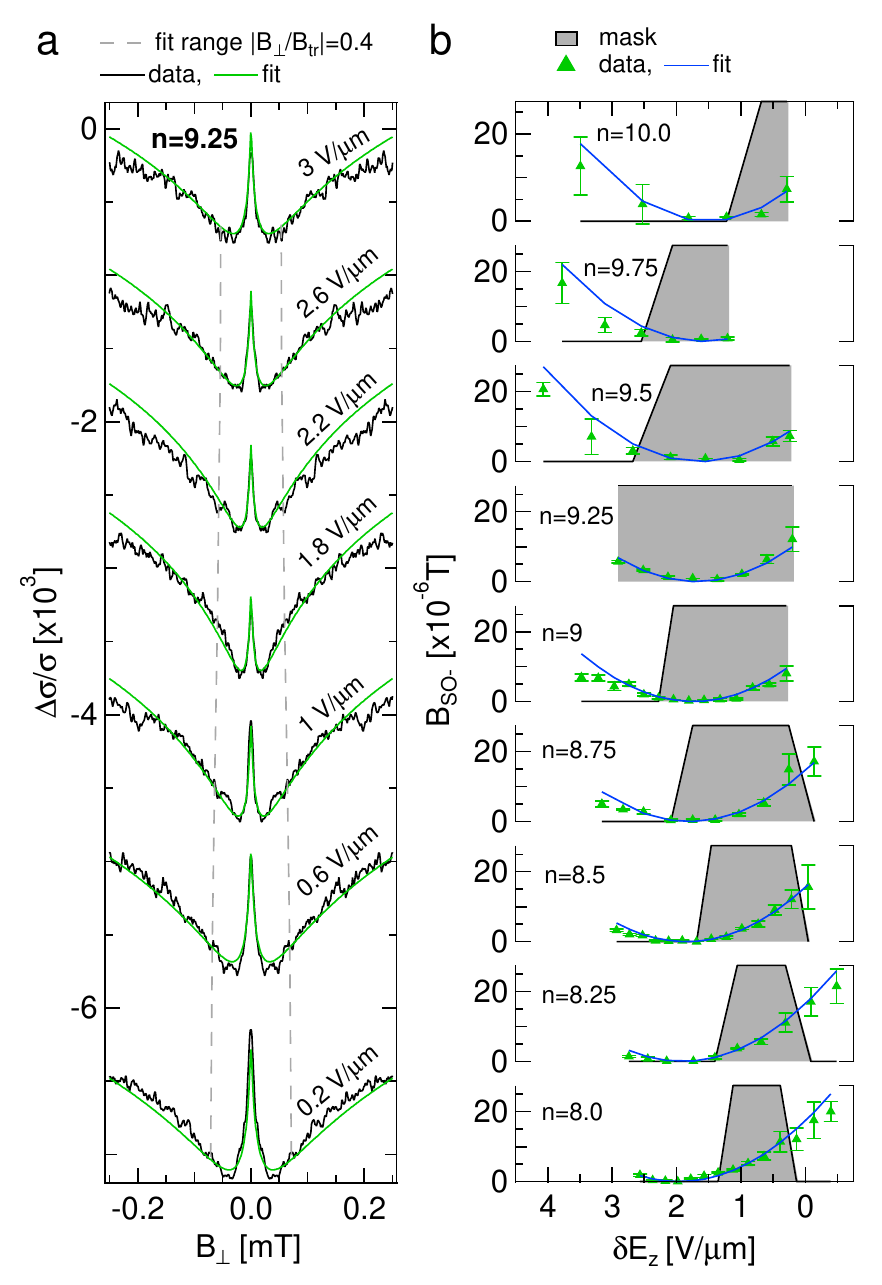}
\vspace{-6mm}\caption{Measured traces away from $\alpha\approx\beta$. (a) Fits (green) to the conductivity traces (black) for constant density $n=9.25\times10^{15}$m$^{-2}$ in \fig{fig:1}. The gray dashed lines indicate the range for the diffusive approximation. Each curve is labeled with its detuning value $\delta E_z$. (b) Extracted values of $B_{\mathrm{SO}-}$ versus the detuning $\delta E_z$ for all densities (arranged vertically and labeled in units of 10$^{15}$m$^{-2}$ for each $B_{\mathrm{SO}-}$ curve). The error bars correspond to the error on the fit parameter (i.e. one standard deviation). The data in the gray shaded area is included in the fit.}
\label{fig:3}\vspace{-6mm}
\end{figure}

We now proceed with the evaluation away from the PSH symmetry by keeping $B_{\mathrm{SO3}}$ and $\tau_\varphi$ fixed for each density, thus facilitating the extraction of $B_{\mathrm{SO}-}$ as a function of the detuning $\delta E_z$. In \fig{fig:3}(a) we show the fits (green) to the conductivity traces along constant density, finding good agreement of the fit with the data. We repeat this for all densities with the respective values of $\tau_{\varphi}$ and $B_{\mathrm{SO3}}$ as previously determined. This delivers a full data set of $B_{\mathrm{SO}-}$ as a function of the density $n$ and the detuning $\delta E_z$. Rewriting \equ{eq:bso2} with the expressions of $\alpha$ and $\beta$ (see \equ{eq:alpha} and \equ{eq:beta}) we obtain
\begin{equation}
B_{\mathrm{SO}-} \propto \left(A+\alpha_1\delta E_z+\frac{1}{2}\pi \gamma n\right)^2,
\label{eq:bsofit}
\end{equation}
with the fit parameters $\alpha_1$ and $\gamma$ and $A=\alpha_0-\gamma \langle k_z^2\rangle$. Thus, the extracted values of $B_{\mathrm{SO}-}$ are expected to follow a parabolic shape, which is also seen in \fig{fig:3}(b). Some deviations from a parabola are apparent, which are due to the non-linear dependence of the density on gate voltages (see \fig{fig:1}). We exclude such data from the fit. The gray shaded area indicates the data points included in the fit -- the fit mask -- considering the validity of the theory and using only the linear region of gate voltage parameter space, see \fig{fig:1} and Appendix Sec. E. The non linear behavior can be seen for larger detunings as the effect of $\delta E_z$ weakens and the $B_{\mathrm{SO}-}$ parabolas become stretched.

The resulting fit to the data is shown in \fig{fig:3}(b) (blue curve), in good agreement with the data within the fit mask and directly yields $A$, $\alpha_1$ and $\gamma$. Self consistent simulations give a value for $\langle k_z^2 \rangle$~\cite{Dettwiler2017}, allowing us to determine the Rashba offset parameter $\alpha_0$ from $A$.

\subsection{Determination of the SO parameters}

\begin{figure}
\includegraphics[width=8.6cm]{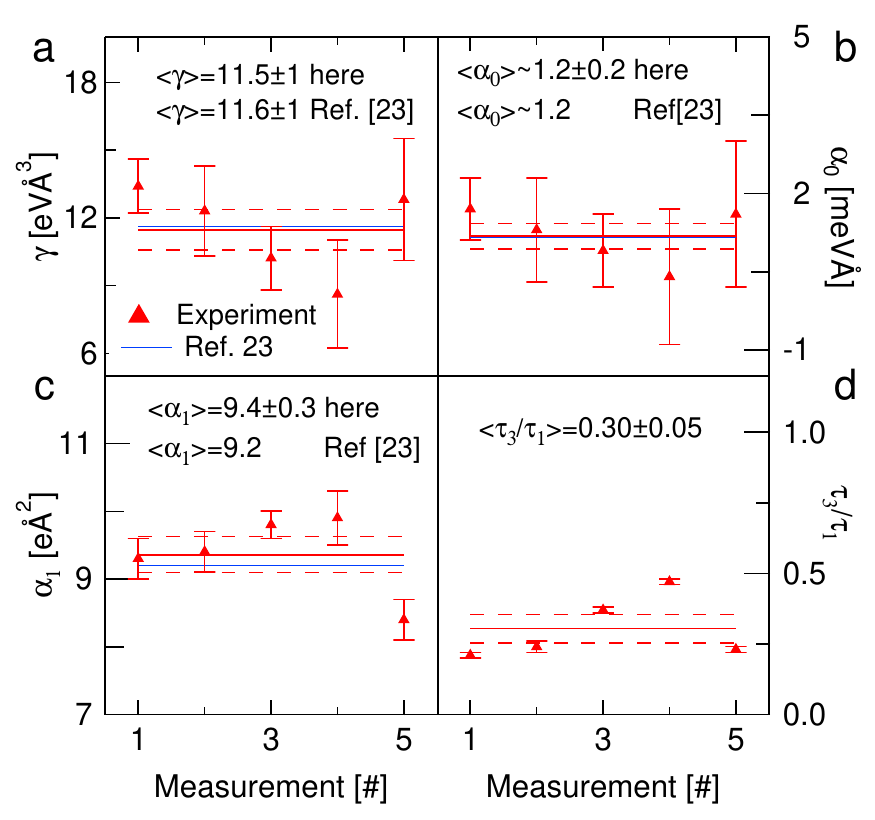}
\vspace{-6mm}\caption{SO parameters from 5 measurements obtained on 2 samples. Measurements 1 and 3 correspond to Hall bar no. I and measurements 2,4 and 5 correspond to Hall bar no. II, where measurement 5 is from another cool down, details in supplementary Sect. II. The blue lines correspond to the values obtained in a previous work~\cite{Dettwiler2017}, the red lines correspond to their average and the red dashed lines correspond to the standard deviation of the mean. (a) Dresselhaus coefficient $\gamma$ (b) offset $\alpha_0$ of the Rashba parameter, (c) $\alpha_1$ of the Rashba parameter and (d) average scattering time ratio $\tau_3/\tau_1$.}
\label{fig:4}\vspace{-3mm}
\end{figure}

In \fig{fig:4} we show the results from 5 independent measurements obtained from 2 Hall bar samples on the same quantum well material (see supplementary Sect.~II). Panels (a) through (c) show the fitted values for $\gamma$, $\alpha_0$ and $\alpha_1$, with their average (red lines) and standard deviation of the mean (red dashed lines). Data sets vary in exact position and especially in number of points measured per density, resulting in varying fit values and associated error bars. To work from the largest possible set of data available we simply include all these independent measurements in the analysis. The complete data sets can be seen on display in the supplementary. The blue lines correspond to the respective values obtained recently from the same wafer material in a previous study~\cite{Dettwiler2017}.

We note that the reported values of $\gamma$ in quantum transport over the last 30 years ranged from $\sim$ 4-28 eV\AA$^3$~\cite{Faniel2011,Miller2003}. The values of $\gamma=28$\,eV\AA$^{3}$ are close to the literature value, which is obtained from $\mathbf{k}\cdot\mathbf{p}$ calculations. However, electronic bandstructure calculations in $\mathbf{k}\cdot\mathbf{p}$ approximation or with density functional theory tend to give inaccurate SO parameters because these calculations neglect either the many body interactions or contain too many parameters which have to be assumed. In recent years, self consistent numerical calculations including the cubic Dresselhaus term were combined with experiments \cite{Krich2007,Dettwiler2017}, giving  values $\gamma \sim$ 9-11.5~eV\AA$^3$. These results are confirmed by state of the art single particle \textit{GW} approximations, calculating the self-energy of a many body system of electrons \cite{Chantis2006} or density functional theory with density dependent exchange potentials \cite{Gmitra2016}. These results agree very well with our average of 11.5\,$\pm$\,1\,eV\AA$^3$ and also recent works using optical spin excitation \cite{Walser2012,Walser2012a,English2013}.

The offset parameter $\alpha_0$ accounts for SO coupling from the electric fields of the charges in the doping layer and the potential of the Hartee term and is a sample specific parameter. It can be calculated via self-consistent methods~\cite{Dettwiler2017,Calsaverini2008}, which is identical with the average of the extracted value. Finally, the Rashba field parameter  $\alpha_1$ has an average value of around 9.4\,e\AA$^2$ which can also be calculated purely from band structure parameters in a quantum well \cite{Calsaverini2008} giving 9.2\,e\AA$^2$, very close to previously extracted values \cite{Dettwiler2017} and ours.

With the previously determined values of $B_{\mathrm{SO3}}$, we can now extract the value of $\tau_3/\tau_1$ using \equ{eq:bso3} and the now known value of $\gamma$. Assuming $\tau_3/\tau_1$ being constant over the range of measured densities, allows us to extract $\tau_3/\tau_1$ from the quadratic fit to the $B_{\mathrm{SO3}}$ data, shown in the upper panel of \fig{fig:2}(b). The fit parameter is proportional to $\gamma\tau_3/\tau_1$ and turns out to be almost the same for all measurements and yields the values shown in \fig{fig:4}(d) by supplying the respective value of $\gamma$ from each measurement. Since $\tau_3/\tau_1\propto1/\gamma^2$, smaller values of $\gamma$ yield a larger $\tau_3/\tau_1$, see data points \#3 and \#4 in \fig{fig:4}(d). From the $B_{\mathrm{SO3}}(n)$ data we can also determine $\tau_3/\tau_1$ as a function of density $n$, using the extracted $\gamma$, which is shown in the middle panel of \fig{fig:2}(b). The values barely change over the range of measured densities and its average value of $\sim$\,0.2 agrees with the one extracted from the fit to $B_{\mathrm{SO3}}$. Overall, the extracted values of $\tau_3/\tau_1$ are around 0.3, much smaller than 1, indicating that small angle scattering dominates~\cite{Knap1996}.

\section{Conclusion}
We derived a closed-form expression for the quantum corrections in the vicinity of the PSH symmetry, which includes the Rashba and linear Dresselhaus terms, as well as the cubic Dresselhaus term. In transport experiments, we studied how breaking of the PSH symmetry, due to the cubic Dresselhaus term and the deviation of the balanced condition of the linear terms, allows to fully quantify the SO strength in a GaAs QW. We achieved this by carefully identifying the different PSH symmetry breaking mechanisms using quantum interference effects.

From the extracted SO terms we directly obtain fundamental SO parameters such as the Dresselhaus coefficient $\gamma$ and the Rashba parameter $\alpha_1$, which are in good agreement with recent calculations and experiments. Supplying the variance $\langle k_z^2 \rangle$ from self-consistent simulations allowed to determine the offset $\alpha_0$ of the Rashba parameter.

The good agreement of the extracted SO parameters with recent theories is an excellent indicator that the new model accurately describes the quantum corrections in the vicinity of the PSH symmetry and can be used as a tool in future studies, whenever Rashba and Dresselhaus SO strengths are comparable. The capability to extract all relevant SO parameters from quantum transport experiments -- obtained from fits to a new closed-form theory -- opens the door to engineer and control the SO interaction as a useful resource in novel quantum materials such as tailored spin textures, Majorana fermions and parafermions. Further, it can be used to coherently manipulate spins in emerging quantum technologies such as spintronics and quantum computation. This technique is also applicable in other materials where the symmetry-broken PSH regime is accessible.

\acknowledgements
We would like to thank Silas Hoffman, Makoto Kohda, Klaus Richter and Gian Salis for valuable inputs and stimulating discussions, and Michael Steinacher for technical support. This work was supported by the Swiss Nanoscience Institute (SNI), NCCR QSIT, Swiss NSF, ERC starting grant (DMZ), the European Microkelvin Platform (EMP), U.S. NSF DMR-1306300 and NSF MRSEC DMR-1420709 and ONR N00014-15-1-2369, Brazilian grants FAPESP (SPRINT program), CNPq, PRP/USP (Q-NANO), and natural science foundation of China (Grant No.~11004120).

\section{Author Contributions}
P. J. W., J. C. E. and D. M. Z. designed the experiment and analyzed the data. All authors discussed the results and commented on the manuscript. S. M. and D. D. A. designed, simulated and carried out the molecular beam epitaxy growth of the heterostructure. P. J. W. carried out the measurements, D. C. M. derived the new expression for the quantum corrections, F. D. fabricated the samples, J. F. and J. C. E. developed and carried out the simulations and theoretical work.

\section{Appendix: Materials and Methods}

\subsection{Formalism to calculate quantum corrections}
Here, we highlight the most relevant results from the formalism to calculate the quantum corrections. The full procedure to calculate the Cooperon and its eigenvalues is shown in detail in the supplementary materials (SM). Our starting point is the general expression connecting the quantum corrections to the conductivity $\Delta \sigma$ and the Cooperon eigenvalues $C_{i}(\mathbf{q})$,
\begin{equation}
\Delta\sigma = -\frac{2e^{2}D\tau^2_0\nu_0}{\hbar^2}\sum_{\mathbf{q}, i}%
C_{i}(\mathbf{q})\;.\label{eq:cond-gen}
\end{equation}
To determine the relevant singlet and triplet Cooperon modes ($i=0,1,2,3$), we start with the impurity mediated  equation for the Cooperon amplitude ${C}_{\mathbf{p},\mathbf{p}^{\prime}}(\mathbf{q})$
\begin{equation}
\begin{split}
{C}_{\mathbf{p},\mathbf{p}^{\prime}} (\mathbf{q})&= |V_{\mathbf{p},\mathbf{p}^{\prime}}|^{2}\\
&+ \sum_{\mathbf{p}^{\prime\prime}}|V_{\mathbf{p},\mathbf{p}^{\prime\prime}%
}|^{2}G^{+}_{-\mathbf{p}%
^{\prime\prime}+ \mathbf{\hbar q}, \epsilon + \hbar\omega}G^{-}_{\mathbf{p}^{\prime\prime},\epsilon}{C}_{\mathbf{p}%
^{\prime\prime},\mathbf{p}^{\prime}}\;.\label{eq:cooperon}%
\end{split}
\end{equation}
The Cooperon amplitude above represents the effective interaction vertex which renormalizes the impurity scattering potential $V_{\mathbf{p},\mathbf{p}^{\prime}}$. It iteratively includes all higher-order processes (multiple scattering events) involving the scattering off of impurities of two electrons following time-reversed paths described by the retarded and advanced impurity-averaged propagators $G^{\pm}$. We solve Eq.~\ref{eq:cooperon} via an iterative procedure by expanding the Cooperon ampliture in its angular harmonics and in the limit $\hbar q\ll p$ (since $\hbar \mb q = \mb p + \mb p'$ and $\mb p' \approx -\mb p$). After some lengthy but straightforward calculation (SM) we find for the relevant zeroth-order harmonic of the Cooperon amplitude
\begin{equation}
{C}^{(0)}_{\mathbf{p},\mathbf{p}^{\prime}} (\mathbf{q}) = \frac{|V_{\mbf p,\mbf p'}|^2}{\tau_0\mathcal{H}}\;. \label{eq:coop-gen}
\end{equation}
The operator $\mathcal{H}$ in the denominator of the Cooperon is
\begin{eqnarray}
\mathcal{H} & = & Dq^{2} + \frac{1}{\tau_{\varphi}} + D\left\{ \left[
Q_{+}^{2} + Q_{3}^{2}\right] J_{z}^{2} + \left[ Q_{-}^{2} + Q_{3}^{2}\right]
J_{x}^{2} \right. \nonumber\\
& + &\left.  2 Q_{+}q_{z}J_{x}-2Q_{-}q_{x}J_{z}\right\} \;, \label{eq:calh}
\end{eqnarray}
where $J_{x,z}$ are the total spin angular momentum components and
\begin{eqnarray}
Q_{\pm}  &  = &  \frac{2m^\ast}{\hbar^2}(\alpha\pm\beta) \;,\\
Q_{3}  &  = & \frac{2m^\ast}{\hbar^2}\left(\beta_3\sqrt{\frac{\tau_3}{\tau_1}}\right). \label{eq:q}
\end{eqnarray}
We can now diagonalize the Cooperon operator in Eq.~(\ref{eq:coop-gen}), a matrix in the basis of the total angular momentum of the two spins, and obtain the quantum correction from Eq.~(\ref{eq:cond-gen}). In what follows, we carry out this procedure for the case in the presence of a quantizing magnetic field $B_\bot$ relevant for the experimental probing of the weak- (and anti-) localization corrections to the conductivity. As described in detail in the SM, in this case we need to switch to a real space description. This is so because in the presence of a magnetic field we approximate the propagators by simply multiplying their zero-field counterpart by a vector potential ($\mathbf{A}$) dependent phase \cite{Altshuler1980}
\begin{equation}
\tilde{G}^{\pm}(\mathbf{r},\mathbf{r}^{\prime})=e^{\frac{ie}{\hbar}\int_{\mathbf{r}}%
^{\mathbf{r}^{\prime}}\mathbf{A}(\mathbf{l})\cdot d\mathbf{l}}G^{\pm}%
(\mathbf{r},\mathbf{r}^{\prime})\;.
\end{equation}
This standard procedure leads to the change $\mathcal{H}\rightarrow \tilde{\mathcal{H}}$ with
\begin{equation}
\tilde{\mathcal{H}}(\mathbf{r},\mathbf{r}^{\prime}) = e^{i\frac{2e}{\hbar}\int_{\mathbf{r}}%
^{\mathbf{r}^{\prime}}\mathbf{A}(\mathbf{l})\cdot d\mathbf{l}}\mathcal{H}%
(\mathbf{r},\mathbf{r}^{\prime})\;,\label{phase}
\end{equation}
in the denominator of the zeroth-order Cooperon operator; the Fourier transform of $\mathcal{H}
(\mathbf{r},\mathbf{r}^{\prime})$ at zero magnetic field is given by Eq.~(\ref{eq:calh}).

We solve the generalized eigenvalue problem,
\begin{align}
\int e^{ i\frac{2e}{\hbar}\mathbf{A}\cdot(\mathbf{r}^{\prime}-\mathbf{r}%
)}\mathcal{H}(\mathbf{r},\mathbf{r}^{\prime})\psi(\mathbf{r}^{\prime})d\mathbf{r}%
^{\prime}
 =  {\cal E}\psi(\mathbf{r})\;,
\end{align}
with suitable expansions of the integrand in powers of $\Delta\mathbf{r }=\mathbf{r}^{\prime}-\mathbf{r}\ll l$ and define the canonical transformation,
\bea
-i\nabla_{z}  &  =  \sqrt{\frac{2eB_\bot}{\hbar}}\frac{(a-a^{\dag})}{i\sqrt{2}}\;,\\
z + z_0  &  =  \frac{1}{\sqrt{\frac{2eB_\bot}{\hbar}}}\frac{(a+a^{\dag})}{\sqrt{2}}\;, \label{eq:can-0}
\eea
with $z_{0} = k_{x}\hbar/2eB_\bot$ ($k_x$ is the Cooperon wave vector along $\hat{x}$). $a$ and $a^\dag$ are bosonic operators, i.e. $[a,a^\dag] = 1$ that describe the quantization of the Landau levels. We thus obtain the characteristic equation in the number representation,
\bea
 & & \left\{\frac{1}{\tau_\varphi} + D\left(Q_+^2 + Q_3^2\right)J_z^2 + \left(Q_-^2 + Q_3^2\right)J_x^2 \right. \nonumber\\
 & - & DQ_+J_z \sqrt{\frac{4eB_\bot}{\hbar}}(a+a^\dag)
- \left.iDQ_-J_x \sqrt{\frac{4eB_\bot}{\hbar}}(a-a^\dag)\right.\nonumber\\
& + & \left. D\left(\frac{4eB_\bot}{\hbar}\right) \left(a^\dag a + \frac{1}{2}\right)\right\}|u\rangle = {\cal E}|u\rangle \;, \label{eq:mag}
\eea
where $|u\rangle$ is the corresponding eigenket.

In the basis of the total spin angular momentum  associated with the 4-dimensional tensor product of the two spin operators of the electrons in time-reversed path, we evaluate the singlet and triplet Landau eigenvalues $\tilde{\mathcal{E}}_{n, i}={\cal {E}}_{n,i}/ (4DeB_\perp/\hbar)$ of the Cooperon ($i = 0$ corresponds to the singlet state and $i = 1,2,3$ label the triplet state).

The singlet $J = 0, J_z = 0$ solution of the Cooperon equation is immediately factored, as it is diagonal both in the spin and Landau level spaces. With these, the single Cooperon mode generates an eingenvalue for the $n$-th Landau level given by,
\begin{equation}
\tilde{\mathcal{E}}_{n, 0}= n + \frac{1}{2} + \frac{B_\varphi}{B_\bot}\;. \label{eq:n00-dim}
\end{equation}
The remaining triplet equation, from Eq.~(\ref{eq:mag}), is written in the basis of $J = 1, J_z = \{1,0,1\}$ in terms of the effective magnetic fields from
Eqs.~(\ref{eq:bphi})-(\ref{eq:bso3})as
\begin{widetext}
\be
\left|\begin{array}{cccc}
\frac{B_\varphi}{B_\bot} +\frac{B_{\mathrm{SO+}}}{B_\bot} + \frac{B_{\mathrm{SO-}} + 3B_{\mathrm{SO3}}}{2B_\bot}& -i\sqrt{{\frac{B_{\mathrm{SO-}}}{2B_\bot}}}(a-a^\dag)& \frac{B_{\mathrm{SO-}} + B_{\mathrm{SO3}}}{2B_\bot}\\
+ a^\dag a + \frac{1}{2}-\sqrt{{\frac{B_{\mathrm{SO+}}}{B_\bot}}}(a+a^\dag) - \tilde{\cal E}& & \\
& & & \\
-i\sqrt{{\frac{B_{\mathrm{SO-}}}{2B_\bot}}}(a-a^\dag)& \frac{B_\varphi}{B_\bot} + \frac{B_{\mathrm{SO-}} + B_{\mathrm{SO3}}}{B_\bot} + a^\dag a + \frac{1}{2}-\tilde{\cal E}&-i\sqrt{{\frac{B_{\mathrm{SO-}}}{2B_\bot}}}(a-a^\dag)\\
& & & \\
\frac{B_{\mathrm{SO-}} + B_{\mathrm{SO3}}}{2B_\bot}&-i\sqrt{{\frac{B_{\mathrm{SO-}}}{2B_\bot}}}(a-a^\dag)&{\frac{B_\varphi}{B_\bot}}+ \frac{B_{\mathrm{SO+}}}{B_\bot} + \frac{B_{\mathrm{SO-}} + 3B_{\mathrm{SO3}}}{2B_\bot}\\
& & + a^\dag a + \frac{1}{2} +\sqrt{{\frac{B_{\mathrm{SO+}}}{B_\bot}}}(a+a^\dag)-\tilde{\cal E} \end{array}\right|=0\;. \label{eq:mag-t}
\ee
\end{widetext}
 In the limit of $\alpha\approx\beta$, $B_{\mathrm{SO-}}\ll B_{\mathrm{SO+}}$, as well as $B_{\mathrm{SO3}}\ll B_{\mathrm{SO+}}$, leading to a justified cancellation of all off-diagonal terms proportional with $B_{\mathrm{SO-}}$ or $B_{\mathrm{SO-}} + B_{\mathrm{SO3}}$ in \equ{eq:mag-t}. Then, by redefining the canonical transformations to operators $a, a^\dag$ are modified to incorporate the additional translation proportional to $Q_+$,
\bea
-i\nabla_{z}  &  =  \sqrt{\frac{2eB_\bot}{\hbar}}\frac{(a-a^{\dag})}{i\sqrt{2}}\;,\nonumber\\
z + z_0 \mp \frac{\hbar Q_+}{2eB_\bot} &  =  \frac{1}{\sqrt{\frac{2eB_\bot}{\hbar}}}\frac{(a+a^{\dag})}{\sqrt{2}}\;, \label{eq:can-1}
\eea
where $-$ corresponds to $J_z = 1$ and $+$ to $J_z = -1$. Then each mode can be diagonalized independently generating the following triplet eigenvalues,
 \bea
\tilde{\mathcal{E}}_{n, 1} =\tilde{\mathcal{E}}_{n, 2} & = n + \frac{1}{2}  + \frac{B_\varphi}{B_\perp} + \frac{B_{SO-}+3B_{SO3}}{2B_\perp}, \label{eq:sum-mag2a}  \\
\tilde{\mathcal{E}}_{n, 3} & = n + \frac{1}{2} + \frac{B_\varphi}{B_\perp} + \frac{B_{SO-}+B_{SO3}}{B_\perp}\;, \label{eq:sum-mag2b}
\eea
Within the same approximation, the associated eigenstates are written in the tensor product space between the LL and the total angular momentum representations as $|n\rangle \bigotimes |J,J_z\rangle$. Because the modes are obtained from three different canonical transformations, Eq.~(\ref{eq:can-0}) for $J_z = 0$, and Eq.~(\ref{eq:can-1}) for $J_z = \pm 1$, the corresponding orbit center in the position representation is determined by the Cooperon wave vector $k_x$ for $J_z = 0$ and $k_x \mp Q_+$ for $J_z = \pm 1$ respectively. The difference $2Q_+$ between the centers of the parallel-spin Cooperon configurations corresponds to the $Q_+$ separation between the $k_x$ momenta of the single-particle states associated with the $\alpha = \beta$ regime \cite{Bernevig2006}. (The Cooperon has a charge $2e$ vs. the single particle states of charge $e$, hence the halving of the momentum translation along $\hat{x}$.)

 Phenomenologically, this situation corresponds to a decreased coupling between the triplet modes within the same Landau level as the scattering processes do not involve any spin-flipping. The original orientation of the incident particle is preserved as the electron population becomes polarized by the effective field $B_{\mathrm{SO+}}$ along the $\hat{z}$ axis.

After angular integration, Eq.~(\ref{eq:cond-gen}) is properly modified to account for the magnetic field, i.e., $\frac{1}{2\pi}\int q dq \rightarrow\frac{1}{4\pi}\frac {4eB_\perp}{\hbar} \sum_{n}$, and the quantum corrections to the conductivity $\Delta \sigma(B_\perp)$  in the presence of a magnetic field are obtained,
\begin{equation}
\begin{split}
\Delta \sigma (B_\perp) &\sim  \sum_{n=0}^{n_m}\sum_{i = 0,3} \frac{1}{\tilde{\mathcal{E}}_{n, i}} \\
&= \sum_{n=0}^{n_m} \left\{\frac{2}{n + \frac{1}{2}  + \frac{B_\varphi}{B_\perp} + \frac{B_{SO-}+3B_{SO3}}{2B_\perp}}\right. \\
&+ \left. \frac{1}{ n + \frac{1}{2} + \frac{B_\varphi}{B_\perp} + \frac{B_{SO-}+B_{SO3}}{B_\perp}} +
\frac{1}{ n + \frac{1}{2} + \frac{B_\varphi}{B_\perp}}\right\}\;. \label{eq:sum-mag2}
\end{split}
\end{equation}
which upon further manipulations (SM) leads to Eq.~(\ref{eq:dsigma}) in the main text. This is the main theoretical result of our work and essential for the two-stage fitting procedure used to accurately determine all the spin-orbit couplings presented here. We emphasize that the closed form expression for $\Delta \sigma (B_\bot)$ in Eq.~(\ref{eq:dsigma}) contains not only the Rashba, but also the linear and cubic Dresselhaus terms.

\subsection{GaAs Quantum Well Materials}
The sample is a modulation-doped 11\,nm thick GaAs/AlGaAs quantum well, grown by molecular beam epitaxy on a (001) n-doped substrate with two symmetrically placed $\delta$ doping layers, each set back 12\,nm from the quantum well. The highly n-doped substrate serves as a back gate by incorporating a 600\,nm thick low temperature grown GaAs barrier, which pins the Fermi level midgap~\cite{Maranowski1995}. This reduces the effective distance $d_B$ from the QW to the back gate and increases the available range of gate voltages. Using wet etching, two identical Hall bars were defined with a Ti/Au gate of 300\,$\times$\,100\,$\mu m^2$ on top. The 2DEG is contacted with thermally annealed low resistance GeAu/Pt contacts. The annealing parameters were carefully determined to achieve decent contact to the 2D gas without short circuiting the back gate. The top and back gate architecture allows us to keep the density in the QW constant, while changing the electric field detuning $\delta E_z$, which can be calculated in terms of the distances effective $d_T$ and $d_B$ and gate voltages $V_{T}$ and $V_{B}$ of the top- and back gate, using a simple plate capacitor model. The detuning then reads \cite{Dettwiler2017}:
\begin{equation}
\delta E_z =\frac{1}{2}\left(\frac{V_{T}}{d_T}-\frac{V_{B}}{d_B}\right).
\label{eq:detuning}
\end{equation}
The back gate range is [-3\,,1\,]\,V and [-0.3\,,0.6\,]\,V for the top gate, corresponding to a density range of [3\,,12\,]\,$\times$10$^{15}$m$^{-2}$, and mobility range [2\,,14\,]\,m$^{2}$/Vs. Individual density and mobility maps are shown in the supplementary.

\subsection{Measurement Technique}
We perform the experiments in a $^3$He-$^4$He dilution refrigerator with a base temperature of 20\,mK. We measure in a standard four-wire lock-in configuration with a time constant of 100\,ms and a current bias of 100\,nA, chosen to avoid self-heating, which can reduce the coherent part of the signal. After setting the gate voltages for each gate configuration gates were given 20 minutes to stabilize. To observe a clear WL/WAL signal each trace was measured at least 10-20\,times and averaged.

\subsection{Symmetry Point Determination and Value of $B_{\mathrm{SO3}}$}
To obtain a value of $B_{\mathrm{SO3}}$, the symmetry point (i.e. $\alpha=\beta$) has to be determined first. For this we perform fits to the measured conductivity traces for all gate configurations along constant density, but replace the SO fields in the argument of \equ{eq:dsigma} with $B_{\mathrm{SO}}^\ast\propto (\alpha-\beta)^2 + B_{\mathrm{SO3}}$ and the extracted value of $B_{\mathrm{SO}}^\ast$ will show a minima at $\alpha=\beta$ and we can locate the approximate position of the symmetry point for each density, where we can then estimate the value of $B_{\mathrm{SO}3}$ (see supplementary, Sec. III).

\subsection{Fit Mask}
The fit mask ensures that the data points included are described by \equ{eq:dsigma} and have the correct $\delta E_z$. We exclude data from the gate configurations in the non-linear region of the density map (see \fig{fig:1}), where the contours for $V_{B}\lesssim -1$V, start to bend. This bending corresponds to a change in the effective distance $d_B$ to the back gate, which we use to calculate the detuning $\delta E_z$. We suspect unpinning of the Fermi level to be the reason for this change in $d_B$.  For more positive gate voltages we exclude data from gate configurations, where the fit to the conductivity traces no longer matches the data. This gives a lower bound on the validity of \equ{eq:dsigma} and agrees quite well with the condition $B_{SO-}\ll B_{SO+}$ (see red dashed lines in \fig{fig:1}).



%

\end{document}